\documentstyle[12pt,epsfig]{article}
\def\appendix#1{
  \addtocounter{section}{1}
  \setcounter{equation}{0}
  \renewcommand{\thesection}{\Alph{section}}
 \section*{Appendix \thesection\protect\indent \parbox[t]{11.715cm} {#1}}
  \addcontentsline{toc}{section}{Appendix \thesection\ \ \ #1}
  }

\renewcommand{\thefootnote}{\fnsymbol{footnote}}
\newcommand{\newsection}{
\setcounter{equation}{0}
\section}

\def\bea{\begin{eqnarray}}
\def\eea{\end{eqnarray}}
\def\be{\begin{equation}}
\def\ee{\end{equation}}
\newcommand{\tr}[1]{\:{\rm tr}\,#1}
\newcommand{\Tr}[1]{\:{\rm Tr}\,#1}

\def\e{{\,\rm e}\,}
\def\d{\partial}

\newcommand{\rf}[1]{(\ref{#1})}
\newcommand{\non}{\nonumber \\*}
\def \m {\mu}
\def \n {\nu}

\def\fec{\Phi_{\rm  (e)}{}}
\def\fmc{\Phi_{\rm  (m)}{}}

\def\pem{\psi}
\def\pme{\psi^\dagger}

\def\diag{{\rm diag}}
\def\a{\alpha}
\def\1{{}{\bf 1}}
\def \ci {\cite}

\def \G {\Gamma}

\def \a{\alpha}
\def\a{\alpha}

\def \N {{\cal N}}

\def \diag {{\rm diag}}

\def \s{{\rm t}} 
 
 \def \Tr {{\rm Tr}}
\def \tr {{\rm tr}}
\def \P {\Phi}

\def \ve {V_{\rm eff} }

\def \sym {{$\N=4$ SYM }}

\def \N {{\cal N}}
\def \L {{\Lambda}}

\def \foot {\footnote}
\def \bi{\bibitem}
\def \la {\label}
\def \tr {{\rm tr}}

\def \II {{\cal I}}

\def \el{{\rm  (e)}}
\def \mm{{\rm  (m)}}
\def \ov {\over}
\def \l {\lambda}

\begin{document}

\begin{titlepage}
\begin{flushright}
ITEP-TH-4/99  \\
Imperial/TP/98-99/41  \\
hep-th/9902095\\
\end{flushright}
\vspace{.5cm}

\begin{center}
{\LARGE  Effective potential in non-supersymmetric }
\\[.5cm] 
{\LARGE 
  $SU(N) \times SU(N)$ gauge theory}
\\[.5cm]
{\LARGE and interactions  of type 0  D3-branes }\\[.2cm]
\vspace{1.1cm}
{\large A.A. Tseytlin${}^{{\rm a}}$\footnote{\ Also at Lebedev Physics
Institute,  Moscow} 
 and K. Zarembo${}^{{\rm b,c}}$\footnote{\ E-mail: 
 zarembo@theory.physics.ubc.ca/@itep.ru}}
\\
\vspace{18pt}
${}^{{\rm a\ }}${\it Blackett Laboratory,
Imperial College, London SW7 2BZ, U.K.
}\\
${}^{{\rm b\ }}${\it Department of Physics and Astronomy,
University of British Columbia,}
\\ {\it 6224 Agricultural Road, Vancouver, B.C. V6T 1Z1, Canada}\\ 
${}^{{\rm c\ }}${\it Institute of Theoretical and Experimental Physics, }
\\ {\it B. Cheremushkinskaya 25, 117259 Moscow, Russia}
\end{center}
\vskip 0.6 cm

\begin{abstract}
We study some aspects of short-distance interaction  between  
parallel D3-branes in type 0 string theory  as  described   by
the corresponding  world-volume gauge theory.
We compute the one-loop effective potential 
 in the non-supersymmetric
$SU(N) \times SU(N)$  gauge theory (which is a $Z_2$ projection
of the $U(2N)$  $\N=4$ SYM theory) representing 
dyonic  branes
composed of $N$ electric and $N$ magnetic  D3-branes.
The  branes  of the same type  repel
at short distances,  but  an  electric and  a magnetic brane
attract,  and  the forces between  self-dual branes cancel.
The self-dual configuration
(with  the positions of the
 electric and the magnetic branes, i.e. the 
 diagonal entries of the adjoint scalar fields, 
being the same)
is stable  against separation of one electric or one magnetic brane,
but is unstable  against 
certain modes of separation of several same-type branes.
This instability  should be   suppressed in the large $N$ limit, 
i.e.  should be   irrelevant for the large $N$ CFT interpretation
of the gauge theory suggested in hep-th/9901101.
\end{abstract}

\end{titlepage}
\setcounter{page}{1}
\renewcommand{\thefootnote}{\arabic{footnote}}
\setcounter{footnote}{0}

\newsection{Introduction}
In recent papers \ci{POL,KT,KTT,KTTT}
it was suggested that  a study of D3-branes in  non-supersymmetric
type 0  string theory  may be useful in attempts to  extend the 
string/gravity -- large $N$ gauge theory duality  \ci{MA,IG}
to  non-supersymmetric 
Yang-Mills theories.

Type 0B theory \ci{DH}   has unconstrained  Ramond-Ramond 5-form field
and thus contains both  electric and magnetic D3-branes \ci{BG,KT}
which may be  combined to form dyonic 
branes \ci{KT,KTTT}.
The field theory  of light  string modes  corresponding to parallel  
$N$ electric or $N$ magnetic  D3-branes  
 is a truncation of \sym theory in which all fermions are excluded, i.e.
 is  $U(N)$ gauge theory  coupled to 6 adjoint scalars
\ci{KT}.
It is  asymptotically free, and was argued in \ci{KTT}
to have an IR fixed point at infinite coupling. 

In \ci{KT} the 
string-theory cylinder diagram  
expression  for  the potential  between same-type 
D3-branes was found and its large-distance limit (dominated by lightest
states of the closed-string channel)
was compared with the corresponding interaction potential in 
the effective low-energy gravitational theory.
It was found that  at large distances the branes 
attract because of the  contribution of the 
bulk tachyon field   (but  would 
repel 
 if  the
closed-string 
tachyon is removed from the spectrum).

In the supersymmetric type II theory  the leading non-vanishing 
term in the interaction   potential between (e.g., moving) branes  has
the same behavior
 at short and at large distances \cite{dkps,DT}. 
This property is based on a certain non-renormalization
theorem  (cf. \ci{BK}),  
e.g., the coefficient of the $v^4\ov x^4$ term in the string
expression (which, in general, is expected to be a function 
of $x\ov \sqrt{\a'}$, i.e.  to receive contributions from massive open string
modes) 
turns out to be a constant \ci{dkps}.
Since this non-renormalization is a consequence of supersymmetry, 
there is  no reason to expect 
 a   similar  `small distance -- large distance' relation 
in the
non-supersymmetric type 0  theory  case.

The expression for the  cylinder amplitude \ci{KT} 
 seems to imply  (depending on a regularization 
of short-distance divergence)
 that  
branes of the same type repel at short distances.
The force between electric and magnetic branes 
is the same in  value but opposite in sign \ci{KTTT},
i.e. they attract at short  distances.

As in the case of  type II  theory \ci{dkps}, 
the {\it short} distance ($\Delta x  <\sqrt{ \a'}$) 
interaction 
between    D-branes  should be dominated 
by the  light {\it open}  string modes, i.e. should be described 
by the one-loop effective potential in the  corresponding 
gauge theory.\foot{  It is important here that the open 
string channel does
not contain tachyon \ci{POL,KT,BG}. 
Truncation to massless open string modes 
is possible when the masses  of the stretched strings 
are much  less than  those  of excited  open string modes, 
$\Delta x \ov \a'$ $ < { 1 \ov \sqrt { \a'}}$.}
Indeed, the presence of a  repelling
force  between  two electric (or  two 
magnetic) branes  can be  seen   directly from the corresponding field-theory 
calculation of the  scalar effective  potential  in \ci{zar99}
(assuming that one-loop induced masses are fine-tuned to zero).

A self-dual type 0 D3-brane is found  by 
putting together 
an  electric and a  magnetic D3-brane \ci{KT}.
Since  an open string
connecting these two   branes  is  a {\it fermion}
\ci{BG,KT}, the resulting  
low-energy world-volume field  theory  contains 
not only the massless bosons for each of the branes but also the 
 massless  fermions   \ci{KTTT}.
The perturbative type 0 
string  theory calculation of the interaction
potential between  two such self-dual 
branes gives (just as in the  case of type IIB  D3-branes 
\ci{pol}) the 
 vanishing result 
at all distances \ci{KTTT}.
Moreover, the potential between an electric (or magnetic) brane and a self-dual brane
also vanishes.
 These results  may be interpreted as  being due to 
 cancellations
between the bosonic (electric-electric and magnetic-magnetic) repulsion 
and the fermionic (electric-magnetic) attraction.

 The field theory on
$N$ electric and $N$ magnetic  parallel type 0  D3-branes contains,
in addition to  the 
$U(N) \times U(N)$  gauge field and 6 adjoint scalars,  also 
4 Weyl fermions in the 
$({ N}, \overline { N})$  representation of $U(N)\times U(N)$ and
4 Weyl fermions in the
$(\overline { N}, {N})$  representation  \ci{KTTT}.   
This non-supersymmetric  gauge theory 
may be interpreted  as  a special   $Z_2$ 
projection  of the $U(2N)$ \sym theory \ci{KTTT}:
one is to keep the fields invariant under 
 change of sign of fermions combined  with the  global $U(2N)$ 
gauge transformation
$ X \to \II  X \II^{-1}, \ \  \II= \pmatrix{ I&  0\cr  0 & -I\cr}$ \
($I$ is the $N \times N$ identity matrix).
The $\II$-transformation changes signs of the off-diagonal $N\times N$
blocks of $U(2N)$ matrices, implying that
the resulting $SO(6)$ invariant  
 theory should contain  only  diagonal ($U(N)\times U(N)$)
bosons  and off-diagonal (bifundamental)  fermions.

As was argued  in \ci{KTTT}, in the large $N$ limit
  this $SU(N) \times SU(N)$  
 non-supersymmetric 
gauge theory (with $U(1) \times U(1)$ part  assumed
 to be decoupled) 
 is expected to be a  conformal field 
theory.\foot{ Similar (but  somewhat more complicated, 
 having less global symmetry)
large $N$ conformal  `orbifold' gauge theories were considered 
in \ci{DM,KS,LNV}..}
It was checked  that  the  one-loop gauge coupling beta-function 
indeed vanishes, 
while the 2-loop beta-function  and the planar
parts of the 
 one-loop renormalizations of the  scalar potential and the  Yukawa
 coupling matrix
vanish to the leading order in large $N$.\foot{\ For example, 
when expressed in terms of the 't Hooft coupling $\l= g^2_{\rm YM} N$
the 2-loop RG equation becomes
$ 
{d \ov d \ln \mu}\l  = { b_2 \ov N^2} \l^3$, 
i.e. running is suppressed  in the large $N$, $\l=$ fixed
 limit.}


Our aim  below  is to further explore  some  perturbative  
properties   of this  
  gauge theory.
We shall compute  the one-loop 
effective potential   for the
diagonal scalar
fields representing short-distance interaction between separated
branes. 
The resulting expression    will be a 
  generalization of the 
potential in the purely-electric theory  case  \ci{zar99}:
it  will  have a simple 
structure of the sum of the bosonic (electric-electric and magnetic-magnetic)
  and the  fermionic (electric-magnetic)
  contributions. 
In agreement with the string-theory result \ci{KTTT}
 that the self-dual branes  do not interact, 
we  will find that 
 the potential indeed 
 vanishes  in the case  when the scalar field backgrounds
for the  two $SU(N)$  groups
(i.e. the positions of the electric 
and the magnetic constituents) are  taken  to be equal.

 Having found the expression for the one-loop
effective potential (Sect.2), we shall then analyze  stability 
of the self-dual configuration of branes
corresponding to the same
 diagonal entries of the  two  sets of 6 scalar fields  (Sect.3).
While the self-dual configuration is  
 stable  against separation of one electric or one magnetic brane,
we shall find that it is 
 unstable against  separation of several 
same-type (electric or magnetic) branes.
The instability disappears at finite temperature $T$ (Sect.4):
the stack of coincident 
like-charge branes 
becomes a metastable state,
and   dissociation of a multiply charged brane  
into elementary constituents ceases to be energetically favorable at
some $T=T_c$ (this  is similar to  the finite-temperature
restoration of spontaneously broken symmetry in the Higgs model).
As we shall  discuss   in Sect.5, 
this instability   is likely to be    suppressed in the large $N$ limit, 
i.e. should be 
 irrelevant for the large $N$ CFT interpretation
of the $SU(N) \times SU(N)$  gauge theory suggested in \ci{KTTT}.


\newsection{One-loop   effective potential }
To write down the action of the  $SU(N) \times SU(N)$  
gauge theory
and to compute the corresponding  quantum effective  potential
it is  simplest to  view it  as a reduction  to 4 dimensions
of a non-supersymmetric  ten-dimensional gauge theory (this is indeed the
way how it  originates  from string theory).\foot{  For simplicity, we shall first  assume
that the gauge group is $U(N) \times U(N)$ as it 
directly  follows from the `D-branes in flat space picture' \ci{WWW,KTTT}.
The truncation to the $SU(N) \times SU(N)$ case \ci{KTTT} 
 will  be 
easy to do  in the final one-loop expressions since the bosons
of the two groups do not mix and the contributions 
of both $U(1)$'s 
decouple. The difference between $SU(N)$ and $U(N)$ cases is irrelevant in the large $N$ limit.}  
The latter $D=10$ gauge  theory    is 
the   $(-1)^F \cdot  \II$ 
projection \ci{KTTT}   of the 
${\cal N}=1,\ D=10$ supersymmetric
$U(2N)$  Yang-Mills  theory \ci{GSO}.

The  field-theory content (i.e. the low-energy degrees of freedom 
on parallel $N$ electric and $N$ magnetic 
 D3-branes) is then described by  the  $U(N)\times U(N)$
$D=10$ gauge potentials and ten-dimensional Majorana-Weyl spinors in 
$(N,\bar{N})$ and $(\bar{N},N)$ representations of $U(N)\times U(N)$ 
\ci{KTTT}.
Gauge potentials are embedded in $U(2N)$ diagonally
\be
A_M=\left(
\begin{array}{cc}
A_{{\rm  (e)} M}&0\\
0&A_{{\rm  (m)} M}
\end{array}
\right) ,
\ee
where the $N \times N$  Hermitian matrices 
 $A_{{\rm  (e)} M}$ and $A_{{\rm  (m)} M}$ describe massless modes of open 
strings connecting electric with electric  and magnetic with magnetic  
branes, respectively ($M=0,...,9$).
 All fields
depend only on 4 `parallel' coordinates $x^\mu$ ($\mu=0,1,2,3$). 
The internal components  of the  gauge potentials
\be
A_i=\left(
\begin{array}{cc}
\Phi_{{\rm  (e)} i}&0\\
0&\Phi_{{\rm  (m)} i}
\end{array}
\right) 
\ee
are the adjoint scalars ($i=1,...,6$).

The fermions fill off-diagonal blocks of the  $U(2N)$ matrices
\be
\Psi=\left(
\begin{array}{cc}
0&\pem\\
\pme&0
\end{array}
\right) ,  
\ee
where   $\Psi_{\rm (em)} = \pem$ 
 and $\Psi_{\rm (me)}= \pme$  
correspond to the  massless modes of fermionic 
 strings stretched
between electric and magnetic branes.
They satisfy the $D=10$  chirality constraint: 
$\G_{11} \pem= \pem, \ \G_{11} \pme= \pme$.\ \foot{  In more detail, 
the 10-d real  MW  fermions of $U(2N)$ gauge theory 
 are represented  by $\Psi_I$, \ $\G_{11} \Psi_I = \Psi_I$,
\ $I=1,...,4N^2$. 
To  describe the projection 
one multiplies them by the  Hermitian  $2N\times 2N$ matrix  generators 
$T^I$ of $U(2N)$ and  then sets  the diagonal 
entries to zero. 
The off-diagonal field $\psi$ is thus a complex Weyl spinor.
The  matrix form of the fermionic 
action is obtained by using that $\tr(T^I T^J) = { 1 \ov 2} \delta^{IJ}$.
}
The  4-d  action  written in  the ten-dimensional notation is\foot{  Here the
trace is in the fundamental representation and  the canonical
gauge theory coupling 
is related  to the string coupling by 
$g^2_{\rm YM} = 4 \pi
g_s$.}
 \be
S= { 1 \ov 2 g^2_{{\rm YM}} }   \int d^4x\ \left( \tr F^2_{MN}  + 
2i \pem^\dagger \Gamma^0 \Gamma^MD_M \pem \right) \ , 
\la{act}
\ee
where 
$\Gamma^M$ are the 10-d  Dirac matrices.
The covariant derivative $D_M = \d_M + i [A_M,\  ]$
acts on the fermions $\pem$ 
as follows: 
\be
D_\mu =\d_\mu+i(A_{\rm  (e)}{}_\mu\otimes\1-\1\otimes {A}^T_{\rm  (m)}{}_\mu)\ ,
\ \ \ \ \
D_i=\d_i +i(\Phi_{\rm  (e)}{}_i\otimes\1-\1\otimes 
{\Phi}^T_{\rm  (m)}{}_i)\ . \la{derr}\ee 
Note that  $A_\mu$ and $\Phi_i$ are Hermitian, 
i.e.  $ A^* = A^T, \ \Phi^*=
\Phi^T$.

As in  the  \sym theory,  the  classical scalar potential 
\be \tr \left([\P_{\el i}, \P_{\el j}]^2  +   [\P_{\mm i}, \P_{\mm j}]^2\right)  \ee 
 has a
minimum  at  $[\P_{\el i}, \P_{\el j}]=0, \  [\P_{\mm i}, \P_{\mm j}]=0$.
The 
 classical moduli space of the world-volume theory is  thus 
described by 
constant transverse coordinates of $N$ electric and $N$
 magnetic D3-branes
\be\label{cl}
\fec_i=\diag(y^1_{{\rm (e)}i}, ..., y^N_{{\rm (e)}i})\ ,~~\ \ \ \ \ \   
\fmc_i=\diag(y^1_{{\rm (m)}i}, ..., y^N_{{\rm (m)}i}) \ ,  
\la{claa}
\ee
where $y_i$  (having mass dimension 1)
are related to the string
coordinates $x_i$ by
$y_i = { x_i \ov {\a'}}$.
In the case of the $SU(N) \times SU(N)$ theory
$\P_{\el i}$ and $\P_{\mm i}$ are traceless, i.e., 
\be
\sum_{a=1}^N y^a_{{\rm (e)}i} =0\ , \  \ \ \ \ \ \ \ \
\sum_{a=1}^N y^a_{{\rm (m)}i} =0\ . 
\la{zerr}
\ee

Our aim below  is to compute the one-loop 
effective action $\G  = \int d^4 x  \ve (\Phi)$ 
 in this constant scalar field background, 
i.e. the corresponding  effective potential \ci{CW}.\foot{  A
 similar calculation of the one-loop 
effective potential in \sym theory viewed as a reduction of 
the $D=10$  SYM theory was considered   in the general 
case of constant but non-commuting scalar background
 $[\Phi_i,\Phi_j]\not=0$
  in  \ci{ft83}.}
This  effective  potential  vanishes in \sym theory
where the classical moduli space is not deformed at the quantum level, 
but is non-trivial in  the non-supersymmetric 
`projected' theory \rf{act}.

\begin{figure}[t]
\hspace*{5cm}
\epsfig{file=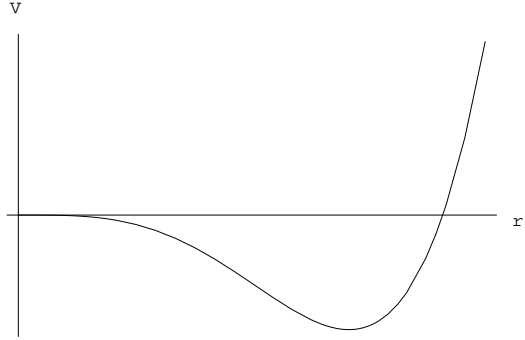,width=7cm}
\caption[x]{Interaction potential \protect\rf{en}
between  two electric  branes.
}
\label{figpot0}
\end{figure}

The bosonic part of the effective potential (which 
describes  interactions of like-charge branes)
 was computed in \cite{zar99}:
\be
V_{\rm eff}^{(bos)}=\frac12\sum_{a,b=1}^N \Bigl[V(|y_{\rm  (e)}^a-y_{\rm  (e)}^b|)+
V(|y_{\rm  (m)}^a-y_{\rm  (m)}^b|)\Bigr] \  .    \la{poten} 
\ee
Here  $|y| = (y_i y_i)^{1/2}$ and the  two-body
interaction potential $V$ is given by the loop integral
\be\label{pot}
V(r)=8\int\frac{d^4p}{(2\pi)^4}\,\ln\left(
p^2+r^2\right)=-\frac{8}{(4\pi)^2}\int_0^\infty\frac{dt}{t^3}\,
\e^{-t r^2}.
\ee
This integral diverges in the UV  region
($t \to 0$)
 and requires renormalization.
We  shall assume that appropriate counterterms cancel quartic
and quadratic
power-like divergencies
and promote the logarithm of the cutoff to the logarithm of
a characteristic
energy scale $\Lambda$ (which should be of order 
of $1\ov \sqrt{\a'}$  in the context of 
 comparison with string theory).
In general, 
$V = c_0\L^4 + c_1 \L^2 r^2 + \frac{1}{4\pi^2}\,r^4\ln\frac{r^2}{\Lambda^2}$.
One may  fine-tune  the renormalization-dependent 
constant $c_1$  (i.e.,  the coefficient of 
the one-loop adjoint-trace  terms 
${\rm Tr}
[ \P_{\el i}^2 +
\P_{\mm i}^2] $)  
to zero
which  would correspond to keeping the scalar fields 
massless at the one-loop  level.
While in the purely electric theory  this 
fine tuning may seem unnatural 
(in the absence of supersymmetry
adjoint scalars may  get masses at loop level \ci{KT}), 
it  is formally possible 
 (see also footnote below  
 eq.~\protect\rf{fros} 
and 
Section 3 
for a discussion of this point).
Fortunately,  this fine-tuning 
 will not be needed 
 in the `self-dual' $SU(N) \times SU(N)$ 
theory 
we are  actually  interested in  where the quadratic 
mass  divergences 
will cancel out (quartic divergences will also cancel out in the  large $N$ limit \ci{KTTT}).
Anticipating that, we shall  assume  the 
following 
   expression for 
the renormalized 2-body potential $V$  \rf{pot} 
\be\label{en}
V(r)=    \frac{1}{4\pi^2}\,r^4\ln\frac{r^2}{\Lambda^2} \ .
\ee
This potential is repulsive at short distances and attractive at
large ones, see fig.~\ref{figpot0} (note that the potential 
becomes attractive at short distances if one adds the mass term $c_1\L^2  r^2$). 
This bosonic part of
the   potential can be compared with the cylinder string
amplitude describing   the interaction between the two  parallel 
electric branes 
 \cite{KT} 
\be\label{sa}
V^{(bos)}_{\rm str} =
-\frac{1}{(4\pi)^2}\int_0^\infty\frac{dt}{2 (2 \pi  \a')^2t^3}\,
\e^{-{  r^2 t\ov 2 \pi  \a'}} \left(\left[\frac{f_3(q)}{f_1(q)}\right]^8
-\left[\frac{f_4(q)}{f_1(q)}\right]^8\right),~~~\ \ \ q=\e^{-\pi t}\ , 
\ee
where we   used the notation 
of  \cite{pol96} (and  suppressed the  space-time volume factor). 
 The short-distance ($ r \ll \sqrt{\a'} $) 
 limit of this expression is obtained 
by expanding 
the integrand for $t \to \infty$ (i.e. $q\to 0 $):\foot{
 Equivalent result is obtained  
taking the limit $x \to 0, \ \a' \to 0,\  y= {r \ov \a'}$ = fixed, 
implying again that $t \to \infty$.}
$f_1(q)=q^{1/12}( 1 - q^2 +\ldots) $, $f_3(q)=q^{-1/24}(1+q+\ldots)$,
$f_4(q)=q^{-1/24}(1-q+\ldots)$. Keeping only the leading term, i.e. 
the contribution of  the massless open-string modes, 
we get (we set $2\pi \a'=1$) 
\be
V^{(bos)}_{\rm str}=- \frac{1}{(4\pi)^2} \
\int_0^\infty\frac{dt}{2t^3}\,
\e^{- r^2 t } \ ( 16 + ... ) \  , \la{fros}
\ee
which is thus the same as
the potential  \rf{pot}  calculated in field theory.
At distances of order $\sqrt{\alpha'}$ 
 the integral over $t$  \rf{sa} 
 is divergent 
at zero ($q \to 1$) 
because of the closed-string tachyon,
 i.e. at string scales the effects 
of the  tachyon condensation \ci{KT}  should  become important.\foot{
 The issue of  UV regularization of the $t$-integral 
is  closely related to that of  the  tachyon
condensation (which is an IR effect from  closed string theory point of view).
The power-like divergencies  would be absent in any
 analytic (e.g., dimensional)  regularization of this  integral which 
 would  ensure that
light  modes of stretched  open strings remain light
 and not get masses of order  $1\ov{\sqrt{\a'}}$.  
Such  a regularization assumption  seems
 to correspond to a special
regime possible  
in the large $N$  limit. This is suggested by  
the existence 
in the   effective  low-energy gravity theory 
of an  electric  large charge 3-brane 
 solution \ci{KT,MIN, KTT}
interpolating between the two $AdS_5 \times S^5$ 
UV and IR 
conformal points. 
 There may be other IR solutions  corresponding to 
 scalars becoming massive \ci{MINA}.} 

\begin{figure}[t]
\hspace*{5cm}
\epsfig{file=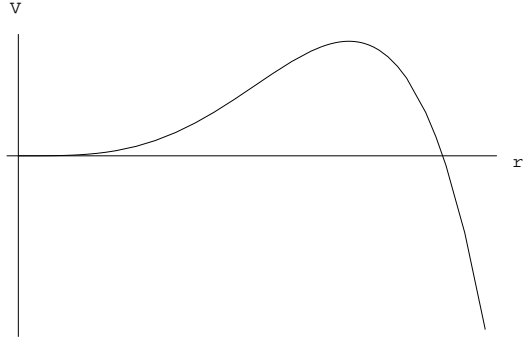,width=7cm}
\caption[x]{Interaction potential \protect\rf{vferm}
between an  electric and a magnetic brane.}
\label{figpot1}
\end{figure}

The contribution of the Weyl  fermion $\psi$ in \rf{act}
 to the effective action is
\be
\Gamma^{(ferm)}=
- \frac12\,\Tr\ln\left[\left(-D^2-\frac{i}{2}\,F_{MN}\Gamma^{MN}
\right)\frac{1+\Gamma_{11}}{2}\right].
\ee
For the  constant 
commuting ($F_{MN}=[D_M,D_N]=0$)  real scalar background \rf{claa} 
one has  
\be
\Gamma^{(ferm)}=-\ 8\ \Tr\ln(-D^2)\ .
\ee
The eigenvalues of $D_\mu=\d_\mu$ and $D_i$ \rf{derr}
 are $ip_\mu$ and  
$i(y_{{\rm (e)} i}^a-y_{{\rm (m)} i}^b)$. The 
fermionic contribution to the effective potential $\ve$ is thus 
\be\label{vferm}
\ve^{(ferm)}=- 8\sum^N_{a,b=1}\int\frac{d^4p}{(2\pi)^4}\,\ln\left[
p^2+ |y_{\rm  (e)}^a-y_{\rm  (m)}^b|^2\right]
=- \sum_{a,b=1}^N V(|y_{\rm  (e)}^a-y_{\rm  (m)}^b|)\ ,
\ee
where the two-body interaction  potential $V$
is the same as in the bosonic
case \rf{pot}.\foot{  We are assuming that 
the bosonic and fermionic 
 determinants are regularized in the same way,
i.e. the scale $\Lambda$ is the same in both cases. 
The string-theory (or \sym projection) origin of the theory
under consideration  implies   that this  is indeed 
  the right regularization 
prescription.  As a result, the
effective potential  will  vanish
 for the self-dual configurations of branes
(see eq.~\protect\rf{int} below), in agreement with the vanishing 
of the   corresponding cylinder 
amplitude in  string theory.}
The  loop  
of the  massless fermionic string  states leads to the interaction between
the electric and the magnetic branes equal 
in magnitude, but opposite in sign, to the bosonic loop  interaction 
between same-type  branes (fig.~\ref{figpot1}).
This is  
in agreement with the full  string-theory result \ci{KT,KTTT} for 
the  electric--magnetic brane interaction  
which is  given by  \rf{sa} taken with the opposite sign.

Collecting together the contributions of the bosonic and the fermionic
degrees of freedom we finally get:
\bea \label{int}
\ve &=&\frac12\sum_{a,b=1}^N \Bigl[V(|y_{\rm  (e)}^a-y_{\rm  (e)}^b|)+
V(|y_{\rm  (m)}^a-y_{\rm  (m)}^b|)-2V(|y_{\rm  (e)}^a-y_{\rm  (m)}^b|)\Bigr]
\non 
&=& { 1 \ov 8 \pi^2} \sum_{a,b=1}^N 
\left[|y_{\rm  (e)}^a-y_{\rm  (e)}^b|^4 \ln { |y_{\rm  (e)}^a-y_{\rm  (e)}^b|^2 \ov \Lambda^2}\  + \  
|y_{\rm  (m)}^a-y_{\rm  (m)}^b|^4 \ln { |y_{\rm  (m)}^a-y_{\rm  (m)}^b|^2 \ov \Lambda^2}
\right. \non &&\left.
\ \ \ \ \ \ \ \ \ \ \ \ \ -\ 2 \ 
|y_{\rm  (e)}^a-y_{\rm  (m)}^b|^4 \ln { |y_{\rm  (e)}^a-y_{\rm  (m)}^b|^2 \ov \Lambda^2} 
\right]\ . 
\eea
In   a generic regularization prescription
for \rf{pot} the  $N^2$ part of the 
coefficient of the $\L^4$ term here 
 cancels out \ci{KTTT}   while  the $\L^2$ term is 
 \be
 (V_{\rm eff})_{\L^2} = \frac{1}{2}\,c_1 \Lambda^2 
 \sum_{a,b=1}^N 
\left[|y_{\rm  (e)}^a-y_{\rm  (e)}^b|^2 \  + \  
|y_{\rm  (m)}^a-y_{\rm  (m)}^b|^2\ -\ 2 \ 
|y_{\rm  (e)}^a-y_{\rm  (m)}^b|^2 \right]\ . 
\la{add}
\ee
Using that $|y_{\rm  (e)}^a-y_{\rm  (e)}^b|^2 
= y_{{\rm  (e)}i}^a y_{{\rm  (e)}i}^a - 2 y_{{\rm  (e)}i}^a
y_{{\rm  (e)}i}^b  + y_{{\rm  (e)}i}^b y_{{\rm  (e)}i}^b$
and \rf{zerr}
it is easy to see that this  combination {\it vanishes}
in  the $SU(N) \times SU(N)$ case, $(V_{\rm eff})_{\L^2}=0$. 
The coefficient of the logarithmic 
divergence in \rf{int} can be transformed with the help of 
\rf{zerr}
 into  the following form
 \be
 (V_{\rm eff})_{\ln \L^2} =-{ 1 \ov 8 \pi^2} 
 \, \ln  \Lambda^2\ \left(
    2 \left[\sum_{a=1}^N  (y_{\el i}^a y_{\el i}^a 
- y_{\mm i}^a y_{\mm i}^a)\right]^2 
+ 4  \left[ \sum_{a=1}^N  (y_{\el i}^a y_{\el j}^a 
- y_{\mm i}^a y_{\mm j}^a) \right]^2     
 \right)\ . 
\la{addi}
\ee
 Note that
 the remaining  dependence on
$\Lambda$  has non-planar  `double-trace' form 
-- the `planar' or  `$N \tr$'    part  has  cancelled out
(${ 1 \ov 8 \pi^2} 
N \sum_{a=1}^N (2 |y_{\rm  (e)}^a|^4 
+ 2 |y_{\rm  (m)}^a|^4 - 2 |y_{\rm  (e)}^a|^4
-  2 |y_{\rm  (m)}^a|^4) =0$) 
in agreement with the general 
conclusion in \ci{KTTT}.\foot{ 
The double-trace term  originating from non-planar diagrams
(with two scalar legs on one boundary of a loop, and two -- on another) 
is  subleading 
in the large $N$ limit -- the traces of the external or 
background
fields are  finite for $N \to \infty$ \ 
 (equivalently, 
one  needs to separate 
a  factor of $N$ in the
 divergent part of the one-loop effective  action 
to combine it with  the gauge coupling  into $N g^2_{\rm YM}$
which is  to be fixed in the large $N$ limit). 
We are grateful to I. Klebanov for a discussion of this issue
and also for suggesting  to put the $\ln \L^2$ 
term in \rf{int}  in the  form   \rf{addi},  which is a special case of the  `double-trace' expression    
 $$ (V_{\rm eff})_{\ln \L^2} =-{ 1 \ov 8 \pi^2} 
 \, \ln  \Lambda^2\ \left(
2 [\tr (\P_{\el i} \P_{\el i})  -\tr ( \P_{\mm i} \P_{\mm i})]^2 
+ 4 [\tr (\P_{\el i} \P_{\el j})  - \tr (\P_{\mm i} \P_{\mm j})]^2 \right)
\ . $$
Although the  counterterm of this form 
when added to the bare action  seems to give the leading 
 order $N^2$ contributions, 
it actually 
produces  only subleading contribution for  $N\rightarrow\infty$ due to
the large $N$ factorization. In calculating diagrams or in the  Schwinger-Dyson
equations one of the traces in the double-trace expression can
be factorized and replaced by its vacuum average which is zero
in the self-dual vacuum. 
Similar remarks apply 
to the $U(N) \times U(N)$ theory (which is equivalent to 
$SU(N) \times SU(N)$ one in the large N limit). In particular, 
the quadratic divergence term  here, while non-vanishing, 
 has (like the logarithmic term) 
the double-trace form, 
$(V_{\rm eff})_{\L^2} =  - c_1 \Lambda^2 
 [\tr \P_{\el i}   -\tr  \P_{\mm i}]^2$,
 and thus  is subleading at large $N$.} 
The remaining finite part of the effective potential 
also originates from non-planar one-loop graphs and 
thus is subleading at large $N$ (see also Sect.5).

We shall discuss some properties of \rf{int}
 in Section 3 and its 
finite-temperature generalization in Section 4.

\newsection{Some properties of  effective potential }
 
There are some  obvious  properties   of the expression \rf{int}
for the effective potential.
First,  the whole potential (and, in particular,  \rf{addi})  vanishes
for a self-dual configuration of D-branes, i.e.  when 
the positions of the electric and the magnetic branes are  taken to 
be the same,  
\be 
y_{\el i}^a=y_{\mm i}^a\ .  
\la{sss}
\ee 
 This is 
 in agreement with the  vanishing of the corresponding   string-theory 
cylinder amplitude 
\ci{KT,KTTT}. 
 It  also  follows from \rf{int}  that  
the  self-dual branes  do not
interact either with electric
or with magnetic branes.


In the case of the  bosonic theory on purely electric branes,
  the 
effective potential 
 \rf{poten} 
\cite{zar99} (defined according to \rf{en}) 
 exhibits the  typical Coleman-Weinberg  behavior \cite{CW} -- 
the $U(N)$ world-volume symmetry appears to be spontaneously broken 
to $[U(1)]^N$ by radiative 
corrections. This means that the stack of  electric  D3-branes 
is unstable against separation;
 in the equilibrium configuration all $N$ branes are
separated
by distances of order $\Lambda$.


The potential of the self-dual brane theory 
 \rf{int} expanded near the self-dual point \rf{sss}
has stable, valley-type, and  unstable directions.
Separating a single electric (or magnetic) brane
away from the rest of  $N-1$  electric and $N$ magnetic branes
gives the same attractive  force 
(for $\Delta y \ll \Lambda$) 
as between a single electric  and a single magnetic brane, 
i.e. this is  a stable direction. An example
of a valley in field space is a separation of 
some number of 
self-dual branes ($y_{\rm  (e)}^s=y_{\rm  (m)}^s $, $s=1,...,M$)
 from the remaining  ($N-M$)  self-dual ones.

If we allow  the electric and magnetic branes to be separated, the effective
potential can become negative due to the repulsion of the same-type 
 branes at
short distances.
This repulsion makes self-dual configuration of branes
{unstable}.
An example of  an unstable direction
is obtained by 
separating 
two
electric branes   along some axis 
by distances $\pm \rho$
from the remaining stack of  coinciding $N-2$ electric
and $N$ magnetic 
branes.
 The energy density  of such configuration is
then the same 
as  of  the  system of  two electric branes at positions $y=\rho$ and $y=-\rho$ 
and two coinciding magnetic branes at the origin $y=0$
 (note that the $N$-dependent contributions  in \rf{int} cancel out)\foot{ This
type of instability depends, of course,  on particular 
 form   of the  two-body interaction potential.
For example, it is absent in the case of the Coulomb-like interaction:
 the forces in the  neutral 
system of two positive charges $q$  at points $y=\pm \rho$ 
and a negative charge  $-2q$   at $y=0$  are attractive
(${ q^2 \ov (2\rho)^2} - 2 { 2 q^2 \ov \rho^2}  < 0$). }
\be
E(\rho) = V_{\rm  eff}  =  \frac{3\rho^4}{\pi^2}\,
\ln\frac{2^{8/3}\rho^2}{\Lambda^2}\ .
\la{unst}
\ee
The self-dual  point $\rho=0$ is a local  maximum, rather than a minimum,
of the energy $E(\rho)$ (see fig.~\ref{figpot0}). 
The branes  thus  tend to separate by
distances
of order $\Lambda$.
 The  behavior of the effective  potential for $\rho \sim\Lambda$
is non-perturbative in  field theory, 
so an equilibrium configuration of branes
cannot be reliably determined in the one-loop approximation.\foot{ Let us note that the electric-magnetic  duality symmetry 
can be interpreted  from the world-volume point of view 
as a $Z_2$ symmetry interchanging $A_\el$  with $A_\mm$,
 $\P_\el$  with $\P_\mm$,
and  $\psi$  with $\psi^\dagger$ in \rf{act}. 
Instability  may  be interpreted as a  spontaneous breaking of 
this $Z_2$ symmetry.}

The above 
 calculation of the effective potential is not
limited
to the case
of equal numbers  of electric and magnetic branes. The world-volume 
theory on a dyonic brane formed by $Q$ electric and $P$ magnetic branes 
has  $U(Q)\times U(P)$ gauge bosons and adjoint scalars 
 and 
fermions in the $(Q,\bar{P})$ and $(\bar Q, P)$
 representations. The effective
potential  in the
generic case is given by the same eq.~\rf{int} with  the appropriate
ranges
of summation ($a=1, ..., Q$ and $b= 1,..., P$).\foot{ 
The
dyonic theory
`interpolates'  between the 
purely electric and the self-dual theory.
 This theory  is not asymptotically
free: its one-loop  gauge theory beta functions 
for the two $U(N)$ couplings 
are proportional to 
${ 8\ov 3}\, (Q-P)$ and ${8\ov 3}\, (P-Q)$, i.e. have opposite signs.}

\newsection{Effective potential at non-zero temperature}
Generalization of the effective potential to the case of non-zero
temperature is straightforward: the  integration
over $p_0$ component of momentum  in \rf{pot}
should be replaced by summation over
Matsubara frequences, even for bosons and odd for fermions. 
The proper-time representation for the bosonic 
 loop integral \rf{pot} at
finite 
temperature  changes to (see, e.g.,  \cite{kap89})
\be\label{pott}
V_B(r,T)=-\frac{8}{(4\pi)^2}\int_0^\infty\frac{dt}{t^3}\,
\e^{- r^2 t }\theta_3\left(\frac{i }{4\pi t T^2}\right).
\ee
The thermal effective potential in the  purely electric theory is thus 
\be
V_{\rm eff}^{(el)}=\frac12\sum_{a,b=1}^N V_B(|y^a-y^b|, T) \ . 
\ee
\begin{figure}[t]
\hspace*{5cm}
\epsfig{file=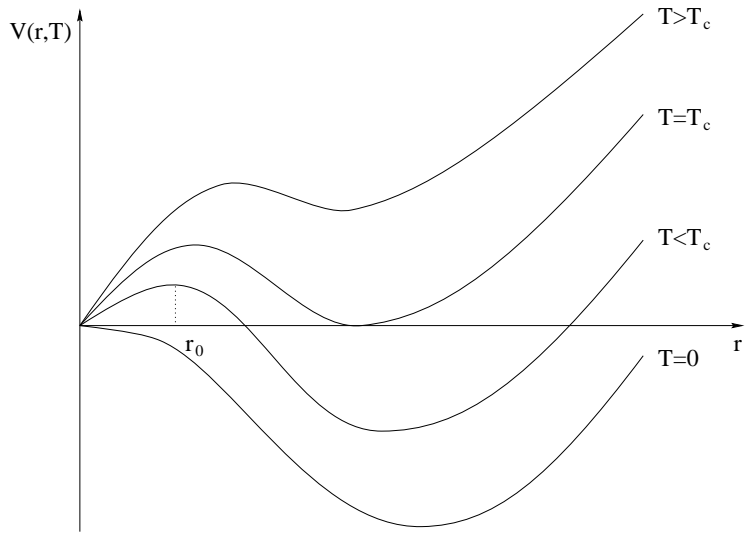,width=7cm,angle=-90}
\caption[x]{Interaction potential between electric branes
 for different temperatures. The potential is normalized so 
that $V(0,T)=0$.}
\label{figpot}
\end{figure}
The temperature qualitatively changes the behavior of the  effective 
potential. At short distances it becomes attractive,
since combining the
high-temperature expansion
of the thermal correction to the effective potential \cite{kap89}
 (the expansion parameter is  $r\ov T$)  with its 
zero-temperature value \rf{en} we get:
\be
V_B(r,T)=-\frac{8\pi^2}{45}\,T^4+\frac23\,T^2r^2-\frac{4}{3\pi}\,Tr^3
-\frac{1}{2\pi^2}\,r^4\ln\frac{\Lambda}{T}+\ldots\ .
\ee
The temperature is assumed to be small compared to $\Lambda$
(i.e. $r < T < \Lambda$) 
 and the
coefficient in front of  $r^4$ is written down with  logarithmic
accuracy.
The region ($0 < r < r_0$) where forces between branes are attractive is very small --
the potential has a maximum (see fig.~\ref{figpot}) at  
\be
r_0=\frac{\pi T}{\sqrt{\frac{2}{3}\,\ln\frac{\Lambda}{T}}}   \   .
\ee
Although at large $r$ the contribution of massive open string states can no
longer be neglected and large-distance interaction of D-branes is not
under control in the world-volume field theory, 
 one  expects, as in \cite{zar99}, that 
 the potential has a global minimum at
some $r\sim\Lambda$. The location of the potential peak, $r_0$,  increases
with the temperature, along with the hight of the potential barrier. One also  
expects that the free energy in the global minimum should  grow \cite{kap89}.
At some temperature $T=T_c\sim\Lambda$, the free energies at zero and at
the symmetry-breaking minimum become equal to each other 
and for $T>T_c$ the global minimum
of the free energy is at $r=0$ (see fig.~\ref{figpot}). This is the well-known
picture of symmetry restoration at finite temperature \cite{KKL}.

In the high-temperature phase the $U(N)$ symmetry is restored,
i.e. the equilibrium configuration corresponds  to all $N$  branes 
 accumulating  at one point.
Similar conclusion is reached in the case of
the type IIB  theory  D3-branes  \ci{ty98}.
The strong-coupling counterpart of this fact 
is non-existence of stable separated-brane supergravity
solutions in the non-extremal case  -- 
when the energy of the system is larger than the total charge,  
branes attract and  form a single black brane.

In the  $U(N) \times U(N)$  theory of 
$N$  electric and $N$ magnetic branes the finite temperature  
effective
potential has  the  same structure 
 as at $T=0$, eq.~\rf{int},
\be\label{intt}
\ve =\frac12\sum_{a,b=1}^N \Bigl[V_B(|y_{\rm  (e)}^a-y_{\rm  (e)}^b|,T)+
V_B(|y_{\rm  (m)}^a-y_{\rm  (m)}^b|, T)-2V_F(|y_{\rm  (e)}^a-y_{\rm  (m)}^b|,T)\Bigr], 
\ee 
but now  the  fermionic and bosonic two-body  potentials
$V_B$ and $V_F$  are not the same.
The proper-time representation for the fermionic loop 
contribution is 
\be\label{potft}
V_F(r,T)=-\frac{8}{(4\pi)^2}\int_0^\infty\frac{dt}{t^3}\,
\e^{- r^2 t }\theta_4\left(\frac{i}{4\pi t T^2}\right)\ , 
\ee
so that at short distances \cite{kap89}
\be
V_F(r,T)=\frac{7\pi^2}{45}\,T^4-\frac13\,T^2r^2
-\frac{1}{2\pi^2}\,r^4\ln\frac{\Lambda}{T}+\ldots \ . 
\ee 
As we have seen in the previous section,  at zero temperature 
 the self-dual 
 vacuum \rf{sss}
 of  the
world-volume gauge 
theory
is unstable 
because of the  repulsive forces between 
same-type  constituent  branes. 
For the same reasons
as in the purely bosonic theory,  we expect   that 
 this instability disappears at 
some critical temperature of order $\Lambda$
 (the self-dual 
vacuum is metastable at any non-zero temperature).    
As in the type II theory case, 
at finite temperature
separated  self-dual  branes 
 start  attracting 
 and should form a single-center cluster. 

\newsection{Discussion }

In this paper we considered some aspects 
of short-distance interactions between 
D3-branes in type 0 string theory 
described by the corresponding  world-volume
field  theory. We concentrated on
the self-dual branes in flat space at weak coupling, i.e. on the
perturbative    $SU(N)
\times SU(N)$  gauge  
theory of \ci{KTTT}.
We  computed  the  one-loop effective potential in this 
field theory  and found 
that the self-dual
D3-brane  configuration \rf{sss} 
 is  perturbatively unstable -- constituent
 electric and magnetic branes tend to separate   
to  distances $\a' \Lambda$ which should naturally be of order 
$ 
\sqrt{\a'}$.

In the large $N$  limit this $SU(N) \times SU(N)$ 
 theory is expected to have  a dual  
representation in terms of 
  the classical type 0 string theory on $AdS_5
\times S^5$
background    with $N$ units of electric and $N$ units of magnetic
5-form flux \ci{KTTT}. 
This implies   that the large $N$ limit of this  gauge theory 
should  represent  a 4-d CFT \ci{KTTT}. 

The perturbative
instability of the self-dual configuration 
 discussed above 
 should not modify this conclusion --
 it is  to be absent in the  part of the effective potential which is 
dominant in the large $N$ 
limit.\foot{ Indeed, it  may seem  natural to ignore the effect of
separation
of  a few  electric or magnetic  branes  from a cluster of large number ($2N$) 
of  branes. Simultaneous separation of a large  number $M \sim N$ 
of branes  should be  statistically suppressed.}

The $SU(N) \times SU(N)$ 
 theory  has a  classical moduli  space  with coordinates  being  
 positions of  the branes  $(y^a_\el, y^a_\mm)$.    
 Together with 
 $ N$ and $g^2_{\rm YM}$   they play the role of  parameters
of the theory.  The  CFT  should be defined  
 by a fixed point in the whole  parameter space: 
$N\to \infty, \ N g^2_{\rm YM} = \lambda,\    
y^a_\el=  y^a_\mm$.
 The  formal one-loop 
  quantum-mechanical instability  of the self-dual point
should   be irrelevant in this context: 
 the large $N$ CFT may  be   
 defined  by a proper  set of conformal composite  operators 
(and their correlation functions)  
which, roughly,   does not include  operators  
vanishing for $\P_\el = \P_\mm$. Like  for the `non-planar' 
  logarithmically divergent terms 
$(\tr \P^2_\el - \tr \P^2_\mm)^2$  in the one-loop 
effective action \rf{addi},  the   contributions 
of such   operators  in correlation functions
 should  vanish 
in the conformal limit. That means that  fluctuations
with $\tr \P^2_\el \not= \tr \P^2_\mm$  should be   effectively forbidden, 
i.e. the  above instability   should be 
 suppressed.\foot{ We are grateful to I. Klebanov
for this suggestion.}

As was recently pointed out \ci{SSN,SN}, the 
non-supersymmetric $SU(N) \times SU(N)$ theory 
interpreted  as a $Z_2$ projection of the  $U(2N)$ $\N=4$ SYM 
theory in  \ci{KTTT}
is also  a special case of $Z_2$  orbifolds
of the  $\N=4$ SYM theory considered in \ci{KS,LNV}
with $Z_2$ here  being in the center of the $R$-symmetry
 group $SU(4)$. 
This implies \ci{SN} that all planar graphs  in this 
theory are the same as in the $U(2N)$  $\N=4$ SYM theory 
restricted
 to invariant  external states.
  In particular, 
 all  planar graph contributions to the scalar effective potential
should vanish, as they do in the  SYM theory
when restricted to the classical or on-shell 
(constant commuting) values of  scalars.\foot{Here one is 
interested in the classical value of the effective potential
(or vacuum energy) and not  in what kind  of scalar  operators  
are induced by loops. For example,  planar and non-planar  1-loop
graphs induce $N \tr(...)$ and $\tr(...) \tr(...)$ 
 operators which produce the same-order $N^2$ contributions
 when  formally inserted back in the loop expansion.}
As a consequence,  the instability discussed above should be absent 
in the  large $N$  theory  to all orders in perturbation theory.



\newsection{Acknowledgements}
We are grateful to H. Liu  and G. Semenoff for discussions
and especially to 
I. Klebanov  for important suggestions 
and   clarifications.
The  work  of A.A.T. was
supported in part
by PPARC, the European
Commission TMR programme grant ERBFMRX-CT96-0045
and  the INTAS grant  96-538.  
The work of K.Z. was supported by NATO Science Fellowship and, in part, by
 INTAS grant 96-0524,
 RFFI grant 97-02-17927
 and grant 96-15-96455 for  the promotion of scientific schools.



\end{document}